\documentclass[epj]{webofc}
\usepackage[caption=false]{subfig}
\begin{document}
%
%\linenumbers
%
\title{Towards an anagraphical picture of high-energy Galactic neutrinos}
%\subtitle{Do you have a subtitle?}

\woctitle{$7^{th}$ Roma International Conference on AstroParticle Physics}

\author{Antonio Marinelli\inst{1}\fnsep\thanks{\email{antonio.marinelli@pi.infn.it}} \and
        Dario Grasso\inst{1} \and
        Sofia Ventura\inst{2}
        % etc.
}

\institute{INFN and Dipartimento di Fisica, Universit\`{a} di Pisa, Largo B. Pontecorvo 3, I-56127 Pisa, Italy
\and
           INFN and Dipartimento di Fisica, Universit\`{a} di Siena, Pian de Mantellini 44, 53100 Siena, Italia
          }

\abstract{%
The TeV/PeV neutrino emission from our Galaxy is related to the distribution of cosmic-ray accelerators, their maximal energy of injection as well as the propagation of injected particles and their interaction with molecular gas. In the last years Interesting upper limits on the diffuse hadronic emission from the whole Galaxy, massive molecular clouds and Fermi Bubbles were set by the IceCube and ANTARES as well as HAWC and Fermi-LAT observations. On the other hand no evidence of Galactic point-like excess has been observed up to now by high-energy neutrino telescopes. This result can be related to the short duration of the  PeV hadronic activity of the sources responsible for the acceleration of primary protons, possibly including supernova remnants. All these aspects will be discussed in this work.}
\maketitle
\section{Introduction}
The astrophysical neutrino flux measured by the IceCube collaboration in the last 8 years gives rise to the need of estimating the Galactic and the extra-Galactic high-energy neutrino fluxes.
While for the second component the debate about the source candidates is still open and more years of multi-messenger observations are needed to provide a solid answer, for the first component we already have a defined picture. Stringent constrains have been in fact obtained on the diffuse Galactic neutrino emission through IceCube and ANTARES analyses and important bounds come from the very-high-energy (VHE) gamma-ray observations of the Fermi Bubbles, the central molecular zone (CMZ) and the possible Galactic Pevatrons. For the diffuse Galactic emission new upper limits (ULs) were obtained with the joint ANTARES/IceCube analysis considering 2780 days of the Mediterranean-based experiment and 2431 days of the South Pole-based experiment~\cite{2018ApJ...868L..20A}. These ULs set the maximal neutrino flux produced by the interaction of Galactic cosmic rays with the gas and consequently also the relative contribution to the full sky astrophysical neutrino flux measured by IceCube. This was possible thanks to a maximum likelihood analysis using the data of neutrino experiments and the map of the neutrino flux expected from the  \emph{Gamma Model}~\cite{2015ApJ...815L..25G}, as discussed in more detail in Sec.~2. Another component that deserve a special focus in our Galaxy is the possible neutrino flux associated with the Fermi Bubbles region~\cite{2014PhRvD..90b3016L,2014ApJ...793...64A}. Fermi-LAT observations from the $\pi$ sr solid angle of the Fermi Bubbles leave space for both leptonic and hadronic interpretations~\cite{2015ApJ...808..107C}. 
However the ULs posed recently by the HAWC observatory on the northern Bubble are compatible with a maximal cutoff in the gamma-ray spectral energy distribution (SED) of few tens of TeV~\cite{2017ApJ...842...85A}.  This is 
showed in more details in this proceeding using the gamma-ray observations collected up to now (see fig.~\ref{Fermi-Bubbles-fig}). An interesting observation of the low latitude Fermi Bubbles SED showed up in 2017 with a new Fermi-LAT analysis~\cite{2017ApJ...840...43A}. The Fermi-Bubbles SED for $\theta \leq 10^{\circ}$\footnote{With $\theta$ representing the Galactic declination.} does not present any cutoff and may be suggesting a different emitting process respect to the outer Bubbles region. In this work we compare the SED obtained by the Fermi-LAT collaboration for the low latitudes region, after the subtraction of point-like and local diffuse contributions, with the SEDs of the expected diffuse Galactic emission. This comparison is reported to verify if a different diffuse Galactic contribution may change the interpretation of the Fermi-LAT data. Other considerations are reported in this proceeding about the possibility to observe Galactic point-like neutrino sources with IceCube and ANTARES using the high energy track-like events.
\label{intro}
\section{News on the Galactic diffuse emission}
\begin{figure}[ht]
\centering
\includegraphics[scale=0.30]{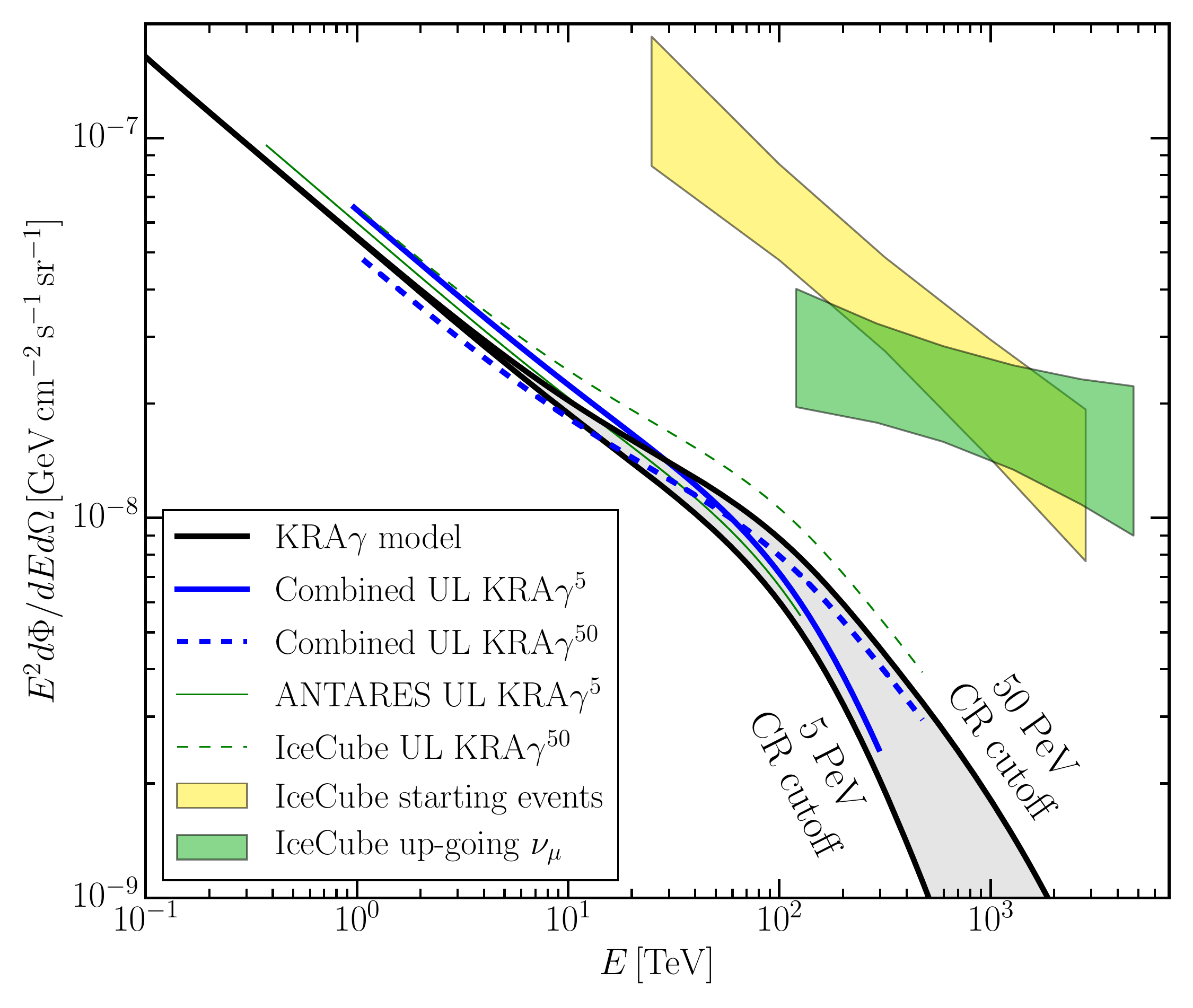}
\caption{Combined ULs at 90\% confidence level  (blue  lines)  on  the  three-flavor  neutrino  flux  of  the KRA$\gamma$
model (\emph{Gamma model}) with  the  5  and  50 PeV  cutoffs  (black  lines) for the primary CRs. The boxes represent the diffuse astrophysical neutrino fluxes measured by IceCube using an isotropic flux template with high energy starting events (HESE) (yellow) and up-going tracks (green)~\cite{2018ApJ...868L..20A,2015ApJ...815L..25G}.}
\label{UL-Gal-ANT-Ice}       % Give a unique label
\end{figure}

The work recently published by ANTARES and IceCube collaborations about the searches for a diffuse Galactic signal gives new interesting ULs about this astrophysical neutrino flux. Using 218 shower-like and 7850 track-like events of ANTARES and 
730130 track-like events of IceCube with a joint analysis it was possible to reach an UL on the diffuse emission due to the Galactic cosmic-ray (CR) sea very close to the \emph{Gamma model} prediction under the assumption of a cutoff  of 5 PeV for CR primary proton energy.  
This limit, which is shown in Fig.~\ref{UL-Gal-ANT-Ice}, amounts to less than 8.5\% of the astrophysical flux measured by IceCube on the whole sky, hence reducing drastically the possibility for the Galactic emission to create a considerable anisotropy in the ``extraterrestrial neutrino'' sky. 
%\label{Galactic}
\section{Fermi Bubbles emission}
\begin{figure}
  \centering
\subfloat[]{{\includegraphics[scale=0.30]{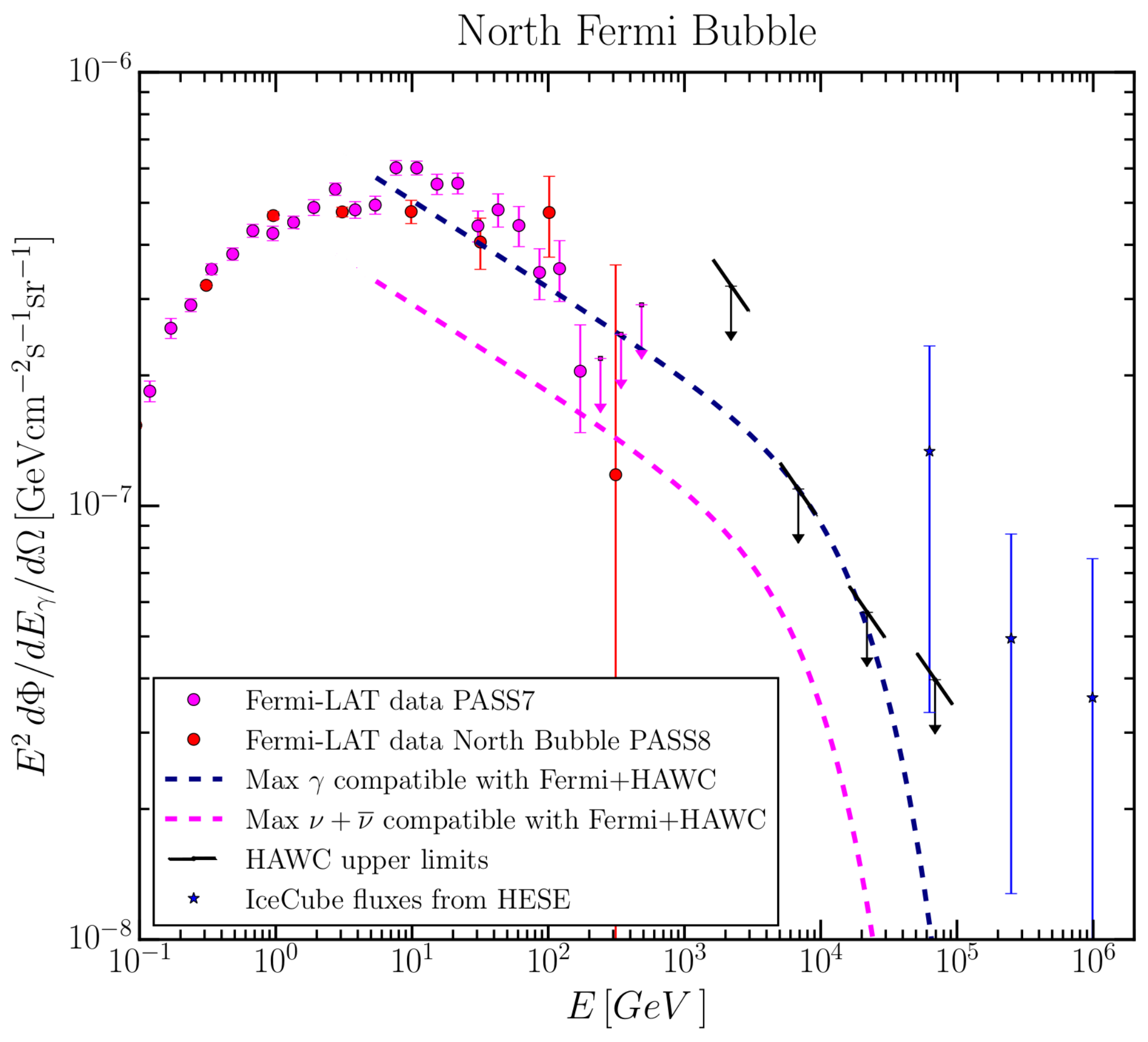} }}%
\qquad
\subfloat[]{{\includegraphics[scale=0.30]{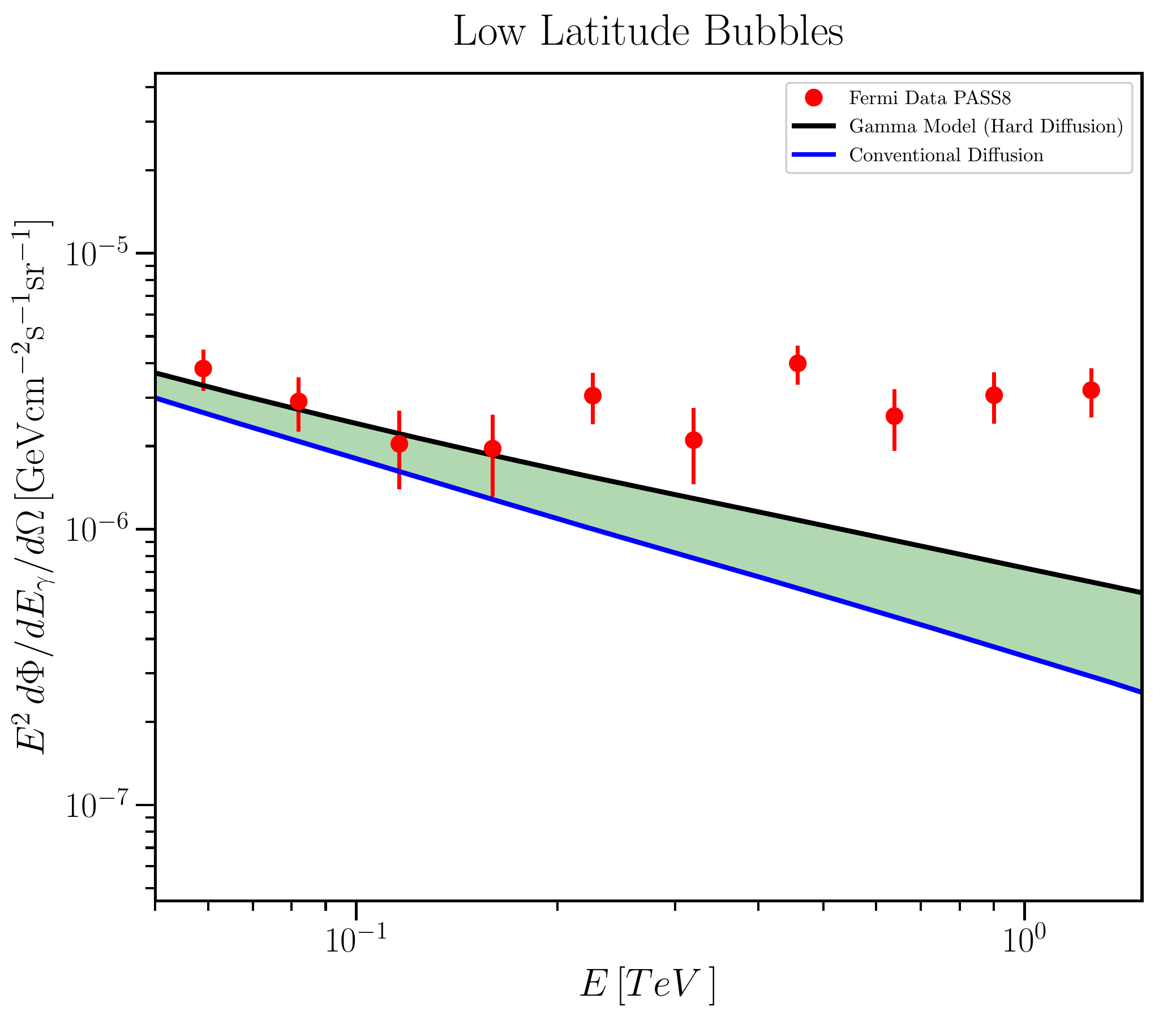} }}%
    \caption{a) SED of the Fermi Bubbles obtained with PASS7 data sample~\cite{2014ApJ...793...64A} and North Fermi Bubble with PASS8 data sample~\cite{2017MNRAS.468.3051N}; the HAWC upper limits~\cite{2017ApJ...842...85A}; the IceCube fluxes limits obtained through the HESE events compatible with the Bubbles; the hadronic fitting for the gamma-ray data, compatible with Fermi-LAT and HAWC observations, with a exponential cutoff at 30 TeV (red dashed line) and the corresponding expected neutrino SED obtained considering the interaction of protons with the photons inside the bubbles (magenta dashed line). b) Fermi-LAT gamma-ray observations for the low latitude Bubbles ($\theta<10^{\circ}$) are shown as a data points while the green region represent the space of gamma-ray SEDs produced in the same region of the sky from the Galactic cosmic rays interaction with the gas, assuming different models. With \emph{Gamma Model} (black line), we intend a harder diffusion due to a radial dependence of the scaling of the diffusion coefficient in addition with a spectral slope change around 300 GeV, as observed by PAMELA, AMS-02 and CREAM~\cite{2015ApJ...815L..25G,2017PhRvL.119c1101G}. Instead, with conventional diffusion (blue line), we intend the uniform CR-sea pervading all the Galaxy.} %
    \label{Fermi-Bubbles-fig}%
\end{figure}

The SED of gamma-ray emission observed by the Fermi-LAT collaboration from the Fermi Bubbles was fitted with a power law with slope $\alpha$ and a exponential cutoff at energy $E_{cut}$. While a hard spectrum ($\alpha\sim 2$) was observed from 1 to 100 GeV a exponential cutoff was set at  $E_{cut}$ of 113 GeV~ \cite{2014ApJ...793...64A}. This SED was described through both hadronic and leptonic emission models, and several studies suggest the possibility to observe high energy neutrinos from this region of the sky assuming that the observed gamma rays are produced by the interaction of protons with photons along the Bubbles. An important constrain to the modeling of the Fermi Bubbles emission at high energies comes from the HAWC observations~\cite{2017ApJ...842...85A}. As showed in Fig.~\ref{Fermi-Bubbles-fig}, the ULs posed by the HAWC observatory on the Northern Fermi Bubble leave space for a possible neutrino SED with a exponential cutoff at around 30 TeV. These ULs reduce considerably the possibility to detect TeV neutrinos from this Galactic source with a maximal corresponding neutrino SED showed in Fig.\ref{Fermi-Bubbles-fig}. Moreover, the IceCube neutrino flux limits obtained considering the IceCube HESE events spatially correlated with this region do not allow a viable connection with the HAWC ULs, as visible in Fig~\ref{Fermi-Bubbles-fig}.
On the other hand new observations of the Fermi Bubbles were performed at low latitudes (previously excluded in the Fermi-LAT analysis), namely the Bubbles region with $\theta<10^{\circ}$. In this new analysis a power-law spectrum without any cutoff was used to describe the observed emission in the Fermi-LAT energy range. This observation suggests a different component in the inner part of the Bubbles and increases the possibility to observe a high energy neutrino signal in case this emission is related to a proton population. However, being this region of the Bubbles coincident with the inner Galactic plane, it is necessary to verify whether this hard SED is contaminated by the enhanced diffuse Galactic component of this part of the Galaxy. In Fig.~\ref{Fermi-Bubbles-fig} we show that the spectrum measured by Fermi-LAT for the low latitude Bubbles presents a spectral component that cannot be reproduced just considering the diffuse Galactic background. This test supports the presence of a different high energy emitting process in the freely-expanding wind zone~\cite{2015ApJ...808..107C}. 
\section{Galactic Pevatrons}
The data collected up to now by IceCube and ANTARES telescopes already set interesting ULs on the potential neutrino emission from Galactic point-like sources~\cite{2017PhRvD..96h2001A,2018arXiv181107979I}, however the possibility to identify hadronic Pevatrons in our Galaxy is matter of debate. Recent studies on Supernova Remnants (SNRs) environment, including calculations on the protons and electrons energy losses and escaping times, show that hadronic Pevatrons can have, at maximum, a duration of few hundreds of years, while the leptonic Pevatrons can last much more~\cite{2018MNRAS.475.5237G}. In other words the VHE emission from a SNR can be dominated by the hadronic emission just during the early stages, whereas we should invoke a leptonic emission to describe the gamma-ray emission observed by the gamma-ray telescopes from the old SNRs. This description of the SNR activity would limit the possibility to observe Galactic point-like neutrino sources. In fact if we consider the angular resolution of $1^{\circ}$ for $\nu_{\mu} + \bar{\nu_{\mu}}$, the time duration of the hadronic PeV emission~\cite{2018MNRAS.475.5237G} plus the diffusion time of protons in the space corresponding to $1^{\circ}$ around the source we obtain at maximum of few thousands of years~\footnote{Using the local B/C for normalizing and a \emph{Gamma Model} diffusion assumption for a Pevatron located in the CMZ we can roughly obtain $t_{diff}(E)= 1.500~years~(L/ 100 pc)^{2} (E / 1 PeV)^{-1/3}$.}. 

\section{Conclusions}
Among all the searches for VHE neutrino emitter candidates, our Galaxy deserves a special focus, with different components that can participate to a potential neutrino flux observable with the IceCube and ANTARES telescopes. In this work we review the possible 
Galactic components of the astrophysical flux measured by the IceCube telescope. We report the recent results obtained thanks to the joint ANTARES/IceCube analysis of the diffuse Galactic emission and highlight the importance of these new constrains. The obtained UL of $8.5\%$ of the total HESE flux (when a cutoff of 5 PeV is selected for the Galactic CR) tell us that the Galactic component can hardly produce a considerable anisotropy in the astrophysical neutrino sky. In this contribution we also compare the diffuse Galactic background with the Fermi-LAT measurements of the low latitude Fermi Bubbles, highlighting the possibility of a new component connected with the freely-expanding wind zone. We show that above 100 GeV the hard SED measured by Fermi-LAT for the low latitude Bubbles cannot be explained with a enhanced diffuse Galactic emission permeating the region. However a contribution of unresolved sources in the Fermi-LAT observed flux cannot be excluded. Finally we point out that the limited duration of the hadronic Pevatron phase of the SNR decreases the possibility to observe a Galactic point-like emission with IceCube and ANTARES telescopes.

\end{document}